\documentclass[twocolumn,secnumarabic,amssymb, nobibnotes, aps, pra]{revtex4-1}

\usepackage{framed}
\usepackage{amsfonts}
\usepackage{amsmath}
\usepackage{graphicx}
\usepackage{color}

\newcommand{\rc}{\textcolor{red}}
\usepackage[hidelinks]{hyperref}
\usepackage{xcolor}
\hypersetup{
	colorlinks,
	linkcolor={blue!80},
	citecolor={blue},
	urlcolor={blue!80!black}
}

\begin{document}
%\maketitle
\title{%Dynamics of Kerr localized states induced by stimulating Raman scattering on the normal dispersion regime
	Influence of Stimulated Raman Scattering on Kerr domain walls and localized structures}
\author{P. Parra-Rivas$^{1,2}$, S. Coulibaly$^{3}$, M. G. Clerc$^{4}$, and M. Tlidi$^5$}

\affiliation{ $^1$OPERA-photonics, Universit\'e libre de Bruxelles, 50 Avenue F. D. Roosevelt, CP 194/5, B-1050 Bruxelles, Belgium\\
 $^2$Laboratory of Dynamics in Biological Systems, KU Leuven Department of Cellular and Molecular Medicine, University of Leuven, B-3000 Leuven, Belgium\\
 $^3$Universit\'e de Lille, CNRS, UMR 8523-PhLAM-Physique des Lasers Atomes et Mol\'ecules, F-59000 Lille, France\\
  $^4$ Departamento de F\'isica and Millennium Institute for Research in Optics, Facultad de Ciencias F\'isicas y Matem\'aticas, Universidad de Chile, Casilla 487-3, Santiago, Chile\\
$^5$Facult\'e des Sciences, Universit\'e Libre de Bruxelles (U.L.B), CP 231, Campus Plaine, B-1050 Bruxelles, Belgium}

\date{\today}


\begin{abstract}
%We study theoretically the influence of stimulating Raman scattering on the formation and stability of localized states in the context of Kerr optical cavities with normal group velocity dispersion. In this regime localized states form due to the locking of domain wall connecting two coexisting continuous wave states, and undergo collapsed snaking. In the presence of stimulating Raman scattering the dynamics, stability and locking of domain walls is strongly modified, what leads to the formation of, not only dark localized states, but also bright ones, which otherwise are absent. We characterize in detail how the bifurcation structure  and stability of these states modifies as a function of the main parameters of the system and classify their dynamics.

%---------------------------------------------
We investigate the influence of the stimulated Raman scattering on the formation of bright and dark localized states in all-fiber resonators subject to a coherent optical injection, when operating in the normal dispersion regime. 
In the absence of the Raman effect, and far from any modulational instability, localized structures form due to the locking of domain walls connecting two coexisting continuous wave states, and undergo a particular bifurcation structure known as collapsed snaking. The stimulated Raman scattering breaks the reflection symmetry of the system, and modifies the dynamics, stability, and locking of domain walls.
% In the presence of stimulated Raman scattering the dynamics, stability, and locking of domain walls is modified.
  This modification leads to the formation of, not only dark, but also bright moving localized states, which otherwise are absent. %Moreover, this effect breaks the reflection symmetry inducing a drift on such states 
We perform a detailed bifurcation analysis of these localized states, and classify their dynamics and stability as a function of the main parameters of the system.

%When this simple device operates in the normal dispersion regime,  the homogeneous (lower and higher) backgrounds of the bistable cycle are both modulational stable, i.e., in a regime far from any
%modulational instability or a Turing type of instability. 

%In this case, the interaction between switching waves or fronts connecting these homogeneous states can stabilize localized states. The Raman effect not only breaks the reflection symmetry due to the causality but also adds a new contribution to the interaction law governing the formation of both dark and bright states. 

% We provide a detailed analysis of bifurcation analysis of these states by drawing their bifurcation diagrams and addressing their stability as a function of the injection strength and the detuning parameter. We show that bifurcation diagrams undergo a collapse snaking type of behavior.  The important issue of our analysis is to reveal the coexistence between moving dark and bright localized states in an all-fiber cavity in a finite range of system parameters. The coexistence between stable dissipative dark and bright localized states in the anomalous dispersion regime is not possible without the stimulating Raman scattering effect.
\end{abstract}
\maketitle

%------------------------------------------------------------------------------------------------------------

\section{Introduction}
Dissipative localized structures (LSs), also known as {\it dissipative solitons}, are coherent states emerging in extended systems far from the thermodynamic equilibrium \cite{nicolis_self-organization_1977,cross_pattern_1993,akhmediev_dissipative_2005,akhmediev2008dissipative,descalzi_localized_2011}. Dissipative LSs may appear in a large variety of pattern forming systems ranging from fluid mechanics and optics, to biology and plant ecology \cite{akhmediev_dissipative_2005,akhmediev2008dissipative,chembo2017theory,tlidi2018dissipative1,tlidi2018dissipative2,malomed2019nonlinear,lugiato2018lugiato,descalzi_localized_2011}. 
%	These emerging structures are often characterized by an intrinsic wavelength which is solely determined by the dynamical parameters, and not by the system size or boundary conditions \cite{prigogine1968symmetry,glansdorff1971thermodynamic,lefever2018rehabilitation}.
%associated with the presence of bi-stability between different coexisting steady states \cite{cross_pattern_1993,akhmediev_dissipative_2005}. In this context, LSs can be seen as a portion of one of those states embedded on the other one.
These robust states can behave like
discrete objects in continuous systems, and can display a variety of different dynamics such as periodic oscillations, chaos, or excitability \cite{akhmediev2008dissipative,descalzi_localized_2011}.
LSs evolve on macroscopic spatial scales, and can be only maintained by permanent non-equilibrium constraints, not being related with the intrinsic inhomogeneities of the
system. Furthermore, once the system parameters are fixed, they are unique, and hence different from the well known conservative solitons that appear as one-parameter
families \cite{akhmediev_dissipative_2005,akhmediev2008dissipative}.
The formation of LSs is usually related with the presence of bi-stability between different coexisting steady states, and therefore LSs can be seen as a portion of one of those states embedded on the other one \cite{coullet_localized_2002}.

%The formation of LSs is normally related to the stable coexistence of two different steady states presence of bi-stability between different coexisting steady states \cite{cross_pattern_1993}. 

Classic examples of dissipative systems where LSs may emerge are found in the field of nonlinear optics and laser physics \cite{akhmediev_dissipative_2005,akhmediev2008dissipative,descalzi_localized_2011}. In this context, LSs have been widely studied in externally driven diffractive nonlinear cavities with 
%either quadratic \cite{etrich_solitary_1997,staliunas_spatial-localized_1998,longhi_localized_1997,oppo_characterization_2001} or
 cubic (i.e., Kerr) nonlinearities \cite{scroggie_pattern_1994,firth_two-dimensional_1996,firth2002dynamical,gomila_excitability_2005}. In these type of cavities, two dimensional LSs, consisting in spots of light embedded on a homogeneous background, form in the transverse plane to the propagation direction, and are commonly known as {\it spatial cavity solitons}.
Similar types of LSs have been shown in wave-guided dispersive Kerr cavities, where they correspond to one-dimensional {\it temporal cavity solitons} emerging along the propagation direction \cite{leo_temporal_2010,leo_dynamics_2013,herr_temporal_2014,xue_mode-locked_2015,garbin_experimental_2017}.
Temporal LSs have been considered as the basis for all-optical buffering \cite{leo_temporal_2010}, and in the last decade, for broadband frequency combs generation in microresonators \cite{delhaye_optical_2007,kippenberg_microresonator-based_2011,pasquazi_micro-combs:_2018}.
In both, diffractive and dispersive Kerr cavities, LSs emerge from a double balance between Kerr nonlinearity and spatial coupling (e.g., diffraction and/or  dispersion) on one hand, and energy gain and  losses on the other hand \cite{akhmediev_dissipative_2005}.
%In the presence of both diffraction and chromatic dispersion, the light can be confined both in space and time leading to the formation of spatio-temporal LSs such as light bullets or light drops \cite{bibid}.

%------------------------------------------------------
%In such type of systems the formation of LSs is usually related with the presence of bistability and heteroclinic tangle between an homogeneous steady state and a subcritical Turing pattern \cite{gomila_bifurcation_2007,leo_temporal_2010,leo_dynamics_2013,parra-rivas_bifurcation_2018}, or due to the locking of domain walls (DWs), switching waves or fronts connecting two different but coexisting homogeneous states.
 
%The existence of optical localized structures either in space and/or in time, due to the occurrence of subcritical modulational or Turing type of instability,
%has been abundantly discussed and is a well-documented issue. Their formation does not require a commutation process between bistable homogeneous steady state, they can be generated in a monostable regime.  However, it is also possible to generate LS far from any modulational or symmetry-breaking instability as a result of interaction between switching waves or fronts in a bistable regime. 

%------------------------------------------------------
\begin{figure}[t]
	\centering
	\includegraphics[scale=0.85]{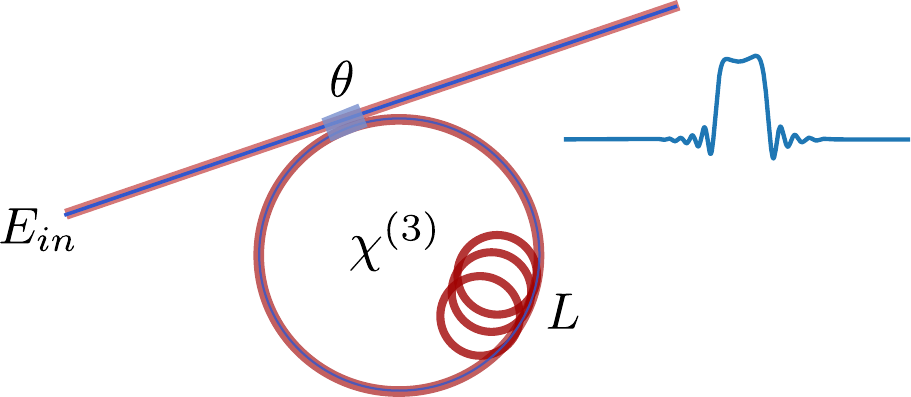}
	\caption{Sketch of an all-fiber cavity of length $L$ driven by an external laser beam of amplitude $E_{in}$ through a beam splitter with intensity transmission coefficient $\theta$. Dissipative structures of different type form and circulate along the cavity. }
	\label{fig0}
\end{figure}

In dispersive Kerr cavities, temporal LSs can form in either the normal or anomalous dispersion regimes.
%In this paper, we focus on the formation of LS in driven dispersive Kerr cavities operating in the normal group velocity dispersion regime.
%In this context LSs form due to the locking of domain walls (DWs),
In the normal regime, LSs arise due to the locking of domain walls (DWs), also known as switching waves or fronts, connecting two different continuous-wave (CW) states \cite{coen_convection_1999,xue_mode-locked_2015, parra-rivas_dark_2016,parra-rivas_origin_2016, garbin_experimental_2017}. These LSs are different from those appearing in the anomalous regime, where their formation is related with the heteroclinic tangle of coexisting CW states and subcritical Turing patterns \cite{gomila_bifurcation_2007,leo_temporal_2010,leo_dynamics_2013,parra-rivas_bifurcation_2018,parra-rivas_dynamics_2014}.
Close to the zero-dispersion wavelength, the influence of high-order effects, such as third and fourth-order dispersion, has to be considered. These terms may cause important modifications on the LSs dynamics, such as the stabilization of different types of LSs, in both normal and anomalous regimes  \cite{tlidi_high-order_2010,tlidi_drift_2013, parra-rivas_third-order_2014, parra-rivas_coexistence_2017,lobanov_dynamics_2017,talla_mbe_existence_2017}. 

In systems made of amorphous materials, such as optical fibers, stimulated Raman scattering (SRS), originated from the delayed material response to electromagnetic excitation, may also have important implications on the LSs dynamics. These implications have been studied by many authors in the context of the anomalous dispersion regime, where most of the studies focus on the cavity soliton dynamics and stability \cite{milian_solitons_2015,chembo_spatiotemporal_2015,lobanov_frequency_2015,yi_theory_2016,wang_stimulated_2018,chen_experimental_2018}. 

In normal dispersion materials, the influence of SRS on the dynamics of DWs and LSs has also attracted an important attention in the last years  \cite{cherenkov_raman-kerr_2017,clerc_time-delayed_2020,yao_generation_2020}. In particular, it has been shown that SRS may stabilize moving bright LSs, which are absent otherwise \cite{clerc_time-delayed_2020}. Close to the nascent bistability onset, where the system can be described by a real order parameter equation, the DWs interaction and locking has been theoretically addressed, and an analytical expression for such interaction has been derived \cite{clerc_time-delayed_2020,chaos_raman}.
However, as far as we known, a complete and detailed description of the bifurcation structure and stability of the LSs in this context is still lacking.
Therefore, the aim of this article is to elucidate the implications that SRS may have, not only on the dynamics and stability of the different type of DWs and LSs arising in these systems, but also in their bifurcation structure.

%---------------------------------------------------
The  paper is organized as follows. In Section~\ref{sec:1}, we introduce the model describing dispersive Kerr cavities in the presence of SRS. Section~\ref{sec:2} focuses on the CW states and the formation of DWs. Later in Sec.~\ref{sec:3}, we introduce the mechanism of DWs locking for the formation of LSs, and we study their bifurcation structure in the absence of SRS. In Sec.~\ref{sec:4}, we analyze the modification of the previous scenario when SRS is considered and how the SRS affects the formation of LSs. Sections~\ref{sec:5} and \ref{sec:6} are devoted to the bifurcation and stability analysis of the Raman LSs. Here we classify the different dynamical regimes in terms of the main parameters of the system. Finally, in Sec.~\ref{sec:7}, a short discussion and the main conclusions of our work are given.

\section{The Lugiato-Lefever model with stimulated Raman scattering}\label{sec:1}

We consider an all-fiber cavity of length $L$ driven by a coherent injected field of amplitude $E_{in}$ as shown in Fig.~\ref{fig0}, where $\theta$ represents the intensity transmission coefficient of the beam splitter. The transmitted part of the injected field circulating within the cavity is affected by Kerr nonlinearity, chromatic dispersion, forcing, and dissipation. In the high-finesse limit, and in the presence of SRS, the intracavity field envelope $E$ of the electric field is described by the extended Lugiato-Lefever equation 
\begin{multline}\label{Eq.1}
t_R\partial_t E=-(\alpha+i\delta_0)E+\sqrt{\theta}E_{in}-i\frac{\beta_2L}{2}\partial^2_\tau E+\\i\gamma L(1-f_R) |E|^2E+if_RL\gamma  E\int_{0}^{\infty}R(\tau')|E(\tau-\tau')|^2d\tau',
\end{multline}
where $\tau$ is the fast time and $t$ is the slow time, $t_R$ is the round-trip time, $\gamma$ is the nonlinear coefficient, $\beta_2$ is the chromatic dispersion coefficient, $\delta_0$ is the phase detuning between the pump field and the nearest cavity resonance, and $\alpha$ represents the linear cavity losses \cite{lugiato_spatial_1987,haelterman_dissipative_1992,chembo_spatiotemporal_2015}. 
The nonlocal delay response term models the SRS, and in agreement with experimental measurements, its kernel or influence function takes the form \cite{lin_raman_2006}
 \begin{figure}[t]
 	\centering
 	\includegraphics[scale=0.95]{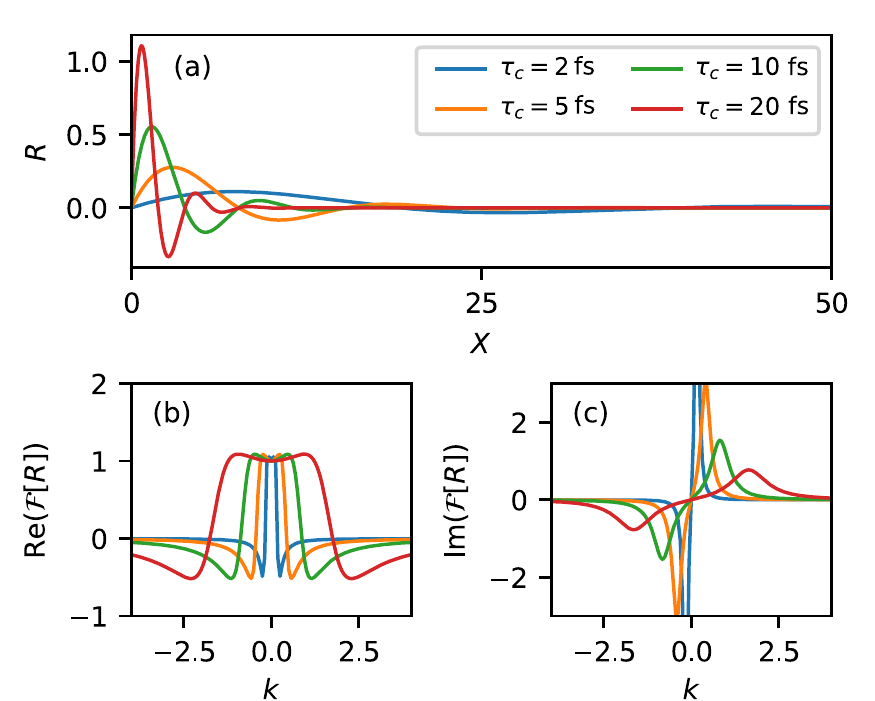}
 	\caption{In (a) the kernel associated with the SRS response function $\mathsf{R}$ for different values of the characteristic parameter $\tau_c$. Panel (b) and (c) shows the real and imaginary parts of $\mathcal{F}[\mathsf{R}]$. Here $(\tau_1,\tau_2,f_R)=(12.2\hspace{0.1cm}{\rm fs},32\hspace{0.1cm}{\rm fs},0.18)$}
 	\label{fig1}
 \end{figure}
 %In agreement with experimental measurements \cite{lin_raman_2006}, the function $R$, or a Kernel describing the nonlocal delayed response should have the following form \cite{agrawal_applications_2008}
\begin{equation}\label{Raman}
R(\tau)=\frac{\tau_1^2+\tau_2^2}{\tau_1\tau_2^2}e^{-\tau/\tau_2}{\rm sin}(\tau/\tau_1),
\end{equation}
where the parameter $f_R$ denotes the strength of the SRS term, and the parameters $\tau_{1,2}$ depend on the type of fiber.
%The factor $(1-f_R)$ in front of the Kerr term is introduced by convenience in order to remove the CW contribution coming from the SRS term.
 Note that in real systems, perturbations due to the higher order dispersion effects may be present alongside the SRS. However, in this theoretical study, for simplicity, we neglect those effects, and focus on the SRS.

The LL equation was first derived to describe passive diffractive cavities \cite{lugiato_spatial_1987}, and later on, in the context of wave-guide dispersive cavities such as fiber cavities  \cite{haelterman_dissipative_1992}, whispering gallery mode resonators \cite{chembo_spatiotemporal_2013}, and integrated ring resonators \cite{kippenberg2011microresonator}. This equation has been also derived in the context of left-handed materials \cite{kockaert2006negative}, for a chain of coupled silver nanoparticles embedded in a glass \cite{ziani2019characterization}, in coupled-waveguide resonators \cite{peschel2004discrete}, and for extended Josephson junctions \cite{cuevas2014sine}. 	
This model constitutes a paradigm for the study of various dynamical properties of laser fields confined in either difractive or dispersive nonlinear optical cavities \cite{chembo2017theory} such as the emergence of patterns \cite{scroggie1994pattern,gomila_transition_2003,gomila_fluctuations_2002,perinet_eckhaus_2017,parra-rivas_bifurcation_2018}, the formation of LS and clusters of them \cite{scroggie1994pattern,gomila_bifurcation_2007,gomila_excitability_2005,godey_stability_2014,parra-rivas_bifurcation_2018,parra-rivas_dynamics_2014,parra-rivas_dark_2016,parra-rivas_interaction_2017,vladimirov_effect_2018}, self-pulsating LS or breathers \cite{firth2002dynamical,leo_dynamics_2013,parra-rivas_dynamics_2014,parra-rivas_dark_2016}, LSs excitability \cite{gomila_excitability_2005}, and some other complex spatiotemporal dynamics such as spatio-temporal chaos  \cite{leo2013dynamics,anderson2016observations,liu2017characterization} or rogue waves \cite{panajotov2017spatiotemporal,panajotov2020control,tlidi2017two,coillet_optical_2014}.

%Furthermore, the polarization properties of periodic patterns and LSs,  
%\cite{hoyuelos1998polarization,averlant2017coexistence,nielsen2019coexistence} and the effect of periodic modulation, either in space or/and time, of the detuning or pump  \cite{de2013phase,odent2014experimental}, have also been analyzed in the framework of this model.

Considering the transformations $E=e_cA$, $\tau=\tau_cX$, and $t=t_cT$,  Eq.~(\ref{Eq.1}) can be written in the dimensionless form 
\begin{multline}\label{Eq.2}
\partial_T A=-(1+i\Delta)A-i\eta_2\partial_X^2A+i(1-f_R)|A|^2A+\\if_RA(\mathsf{R}\otimes|A|^2)+S,
\end{multline}
%where $e_c=\sqrt{\frac{\alpha}{\gamma L}}$, $\tau_c=\sqrt{\frac{L|\beta_2|}{2\alpha}}$, and $t_c=t_R/\alpha$,
where $e_c=\sqrt{\alpha/\gamma L}$, $\tau_c=\sqrt{L|\beta_2|/2\alpha}$, and $t_c=t_R/\alpha$,
and the normalized detuning, pump intensity, and group velocity dispersion coefficients read:
\begin{align*}
\Delta=&\frac{\delta_0}{\alpha},&S=&\sqrt{\frac{\theta\gamma L}{\alpha^3}}E_{in},&\eta_2=&{\rm sign}(\beta_2).
\end{align*}
%In the following we focus on the normal GVD regime and therefore $\eta_2=1.$	
In Eq.~(\ref{Eq.2}), $\otimes$ is the convolution between the intensity $|A|^2$, and the extended Raman kernel $\mathsf{R}$
\begin{equation}
\mathsf{R}(X)\equiv H(X)\cdot R'(X)\equiv H(X)\cdot\eta e^{-a X}{\rm sin}(b X),
\end{equation}
where $H$ is the Heaviside function, and 
\begin{align*}
\eta=&\tau_c\frac{\tau_1^2+\tau_2^2}{\tau_1\tau_2^2},&a=&\tau_c/\tau_2,&b=&\tau_c/\tau_1.
\end{align*}
The SRS term is calculated through the convolution theorem which states
$$\mathsf{R}\otimes|A|^2=\mathcal{F}^{-1}(\mathcal{F}[\mathsf{R}]\cdot\mathcal{F}[|A|^2]),$$
where $\mathcal{F}$ is the Fourier transform. The real and imaginary parts of $\mathcal{F}[\mathsf{R}]$ read
\begin{subequations}
	\begin{equation}
	{\rm Re}(\mathcal{F}[\mathsf{R}])=b\eta(a^2+b^2-k^2)/Z(k),
	\end{equation}
	\begin{equation}
	%{\rm Im}[\mathcal{F}[R]]=\frac{2\eta ab k}{Z(k)},
	{\rm Im}(\mathcal{F}[\mathsf{R}])=2\eta ab k/Z(k),
	\end{equation}
\end{subequations}
with $Z(k)=(a^2+b^2-k^2)^2+4a^2k^2$. The imaginary part of  $\mathcal{F}[\mathsf{R}]$ corresponds to the Raman gain spectrum, while the real one represents the modification of the refractive index due to the SRS term \cite{agrawal_applications_2008}.

%----------------------------------------------------------------------

%----------------------------------------------------------------------

The SRS introduces an additional dependency on $\tau_c$ which becomes an important parameter for controlling the strength of the Raman response function $\mathsf{R}$, and furthermore connects to physical parameters of the cavity such as chromatic dispersion coefficient, length and losses \cite{wang_stimulated_2018,clerc_time-delayed_2020}. The modification of the SRS response with $\tau_c$ is plotted in Fig.~\ref{fig1}(a) for $\tau_c=2,5,10,$ and $20$ fs, and Re$(\mathcal{F}[\mathsf{R}])$ and Im$(\mathcal{F}[\mathsf{R}])$ are shown in Fig.~\ref{fig1}(b) and (c), respectively. While the envelope of $\mathsf{R}$ decreases with $\tau_c$, the wavelength increases [see Fig.~\ref{fig1}(a)], and when $\tau_c$ is very large, (i.e., $\tau_c\rightarrow\infty$), $\mathsf{R}$ becomes very sharp approaching an instantaneous response (i.e., a Dirac delta).

In the absence of SRS, Eq.~(\ref{Eq.2}) is invariant under the transformation $X\rightarrow-X$ (i.e., $X-$reflection symmetry), and its LS solutions normally preserve this symmetry. In contrast, when SRS is taken into account, the reflection symmetry is broken, leading to asymmetric solutions which now drift at a constant velocity $v$, which depends on the SRS control parameters \cite{clerc_time-delayed_2020}.

%	Note that for large values of $\tau_c$  ${\rm Im}(\mathcal{F}[\mathsf{R}])$ flattens [see Fig.~\ref{fig1}] and the system almost conserves the reflection symmetry.

In what follows, we focus on the normal dispersion regime (i.e., $\eta_2=1$), and we fix $f_R=0.18$, $\tau_1=12.2$ fs, and $\tau_2=32$ fs  corresponding to the common parameters of fused-silica based fibers \cite{lin_raman_2006}. With these specifications 
the main control parameters of the system are $\Delta$, $S$ and $\tau_c$. Furthermore, in this work we consider a normalized domain width $l=100$ and periodic boundary conditions.

\section{Continuous wave bi-stability and domain walls dynamics}\label{sec:2}

The stationary states (i.e., $\partial_tA=0$) of this system are described by 
\begin{multline}\label{eq.3}
-(1+i\Delta)A+v\partial_XA-i\eta_2\partial_X^2A+i(1-f_R)|A|^2A\\+if_RA(\mathsf{R}\otimes|A|^2)+S=0,
\end{multline}
where we have considered the comoving frame transformation $X\rightarrow X-vT$, to take into account moving states at constant speed $v$.

The basic state solutions of Eq.~(\ref{eq.3}) are the homogeneous or CW states $A_h$ satisfying
%\begin{multline}\label{eq.3}
%-(1+i\Delta)A_h+i(1-f_R)|A_h|^2A_h\\+if_RA_h(\mathsf{R}\otimes|A_h|^2)+S=0,
%\end{multline}
%which can be further reduced to  
\begin{equation}\label{HSS}
I_h^3-2\Delta I_h^2+(1+\Delta^2)I_h=S^2,
\end{equation}
where $I_h\equiv|A_h|^2.$
\begin{figure}[t]
	\centering
	\includegraphics[scale=0.95]{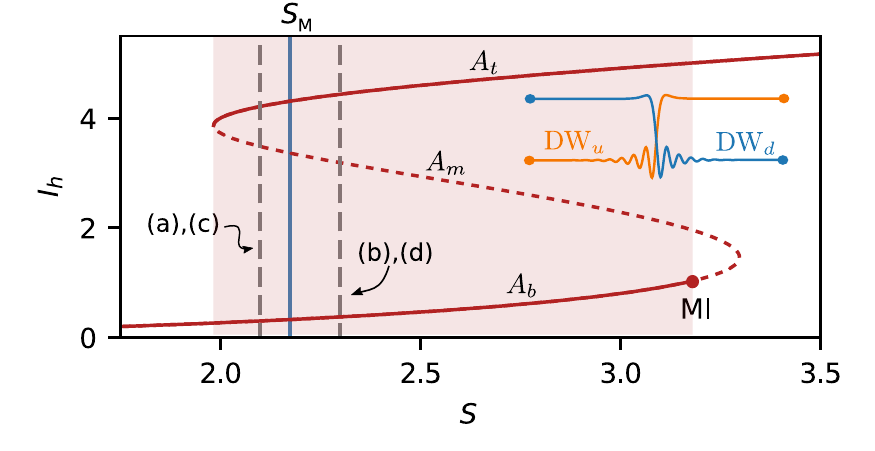}
	\includegraphics[scale=0.95]{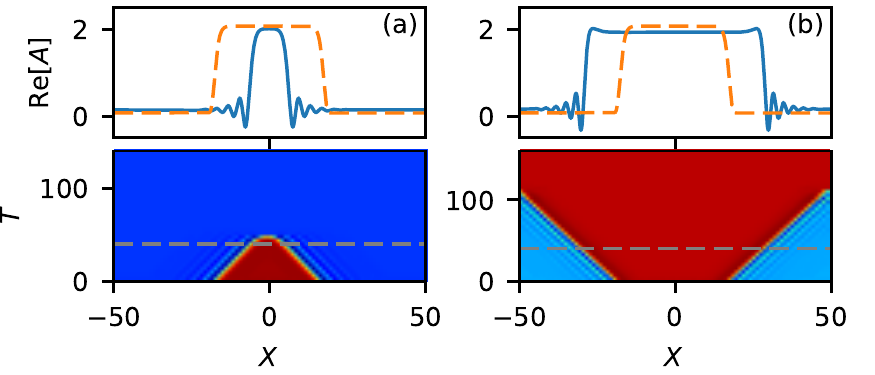}
	\includegraphics[scale=0.95]{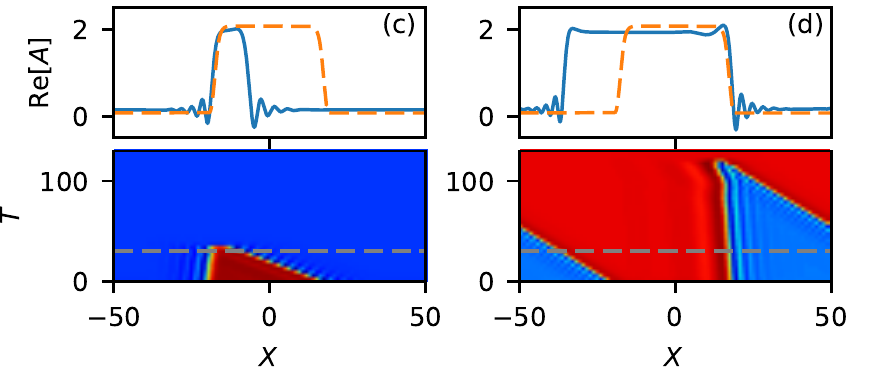}
	\caption{Intensity of the CW state as a function of $S$ for $\Delta=4$. Stable (unstable) branches are shown with solid (dashed) lines. The light red area shows the bi-stability region. The blue vertical line corresponds to the Maxwell point of the system $S_M$, and the two vertical dashed gray lines correspond to the temporal evolutions shown in panels (a)-(d). Panels (a) and (b) show the evolution of two different initial conditions of the form $g^b(X)$ in the absence of SRS. Panels (c) and (d) show the temporal evolution of the same initial condition in the presence of SRS for $\tau_c=5$ fs. The gray horizontal line on the colormaps corresponds to the blue profile on top of them. }
	\label{fig2}
\end{figure}
For $\Delta<\sqrt{3}$ $A_h$ is single-valuate. For $\Delta>\sqrt{3}$, $A_h$ is three-valuate, and therefore compose by the three solutions $A_b,A_m$ and $A_t$, separated by two folds or saddle-node bifurcations SN$_{t,b}$ occurring at 
 \begin{equation}
 I_{t,b}=\frac{2\Delta}{3}\pm\frac{1}{3}\sqrt{\Delta^2-3}.
 \end{equation}
In the later case, CW shows a hysteresis loop like the one plotted in Fig.~\ref{fig2} (top panel) for $\Delta=4$. In the absence of SRS ($f_R=0$), 
%linear stability analysis respect perturbations of the form $e^{\Omega T+ikX}+c.c.$ shows that: 
the middle branch $A_m$ is always spatiotemporally unstable, the top one $A_t$ is always stable, and the bottom one $A_b$ undergoes a modulational instability (MI) at $(S_c,I_c)\equiv(\sqrt{(1+(1-\Delta)^2)},1)$ such that it is stable for $S<S_c$, and unstable between the MI and SN$_b$. We plot stable (unstable) solution branches using solid (dashed) lines. 
In the range between SN$_t$ and MI ($S_c$) (see light red area in Fig.~\ref{fig2}) stable $A_b$ and $A_t$ coexist, and in the following we refer to this interval as the {\it bi-stability} region.

Within the bi-stability region, DWs connecting $A_t$ and $A_b$ either upwards (DW$_u$)  or downwards (DW$_d$) can form (see the inset in Fig.~\ref{fig2}). DW$_u$ and DW$_d$ are related by a reflection $X\rightarrow-X$ respect to their center, i.e., ${\rm DW}_{u}(X)={\rm DW}_d(-X)$, and in general, are not stationary, but move at constant speed and opposite direction depending on the parameters of the system. 

To have some insight about the DWs behavior let us first show how an initial condition of the form
\begin{equation}\label{gauss}
	%g^{b(t)}(X)=A_{b(t)}\pm G(X)=A_{b(t)}\pm h{\rm exp}[X/\sigma]^{10},
		g^{b(t)}(X)=A_{b(t)}\pm h{\rm e}^{-[X/\sigma]^{10}},
\end{equation}
composed by a super-Gaussian profile sitting on $A_h$ evolves in time, with $\sigma$ and $h$ being its standard deviation and height. 
%In the following we fix $h$ to be approximately equal to the separation between $A_b$ and $A_t$, and modify the width of $G(X)$ by controlling $\sigma$.}
%To have some insight about the DWs behavior let us first show how an initial condition of the form
%%g^{b(t)}(X)=A_{b(t)}\pm G(X)=A_{b(t)}\pm h{\rm exp}[X/\sigma]^{10},
%\end{equation}
%evolves with $T$. This initial condition is composed by a super-Gaussian profile of high $h$ which sits on a CW state. In the following we fix $h$ to be approximately equal to the separation between $A_b$ and $A_t$, and modify the width of $G(X)$ by controlling $\sigma$.
%---------------------------------
Figure~\ref{fig2}(a) shows the evolution of Eq.~(\ref{gauss}) for $S=2.1<S_M$, and corresponds to the dashed gray line in the CW diagram. The $g^b$ profile [see orange dashed line in Fig.~\ref{fig2}(a,top)] establishes a connection between $A_b$ and $A_t$, leading to the formation of  DW$_u$ and DW$_d$, which soon after move inwards, with the same speed and opposite propagation direction. Eventually, the DWs annihilate one another,
bringing the system back to $A_b$. A profile along such evolution is shown in Fig.~\ref{fig2}(a,top) for $T=40$.

Figure~\ref{fig2}(b) shows the evolution of the same initial condition for $S=2.3>S_M$. For this value of $S$, the DWs move outwards, and eventually they meet at the boundaries of the domain, where they collide and disappear, such that the system terminates at the $A_t$ state. The transition between these two scenarios takes place at the Maxwell point $S_M$ of the system (see Fig.~\ref{fig2}) where the DWs velocity cancels out \cite{chomaz_absolute_1992,parra-rivas_origin_2016}.

%Figure~\ref{fig2}(b) shows the evolution of the same initial condition for $S=2.3>S_M$. In this case, however, the DWs move outwards, i.e., DW$_u$ moves to the left and DW$_d$ to the right, until they meet at the boundaries of the domain, where they collide and disappear, bringing the system back to $A_t$. The transition between these two scenarios take place at the Maxwell point $S_M$ of the system [see Fig.~\ref{fig2}] where the DWs velocity cancels out \cite{chomaz_absolute_1992,parra-rivas_origin_2016}.

When the SRS is taken into account ($f_R\neq0$), the features and dynamics of DWs change. The SRS breaks the reflection symmetry of the system, and DWs are no more related by the transformation $X\rightarrow-X$ [i.e., ${\rm DW}_{u}(X)\neq{\rm DW}_d(-X)$]. Due to this asymmetry, DW$_u$ and DW$_d$ now move at different speeds and opposite directions, leading to the asymmetric time evolution shown in Figs.~\ref{fig2}(c) and (d). For $S=2.1<S_M$ [see Fig.~\ref{fig2}(c)] DW$_u$ moves slower than DW$_d$, although they eventually collide bringing the system back to $A_b$. For $S=2.3>S_M$ [see Fig.~\ref{fig2}(d)], DW$_u$ moves much faster than DW$_d$, but as before, their annihilation eventually takes place, and the system finally reaches $A_t$.

\section{Formation of localizes states in absence of stimulated Raman scattering}\label{sec:3}

The temporal evolutions shown in Sec.~\ref{sec:2} correspond to values of $S$ far from the Maxwell point of the system, where DWs annihilate one another, leading to one of the CW attractors, either $A_b$ or $A_t$.
Close to the Maxwell point however, DWs may lock at certain separations leading to the formation LSs of different widths \cite{parra-rivas_dark_2016,parra-rivas_origin_2016,clerc_time-delayed_2020}. In this section, 
we illustrate this mechanism in the absence of SRS ($f_R=0$), and later, in Sec.~\ref{sec:4}, we show the implications that the SRS may have on the LSs formation, dynamics and stability. 

\subsection{Locking of domain walls}
\begin{figure}[t]
 	\centering
 	\includegraphics[scale=1]{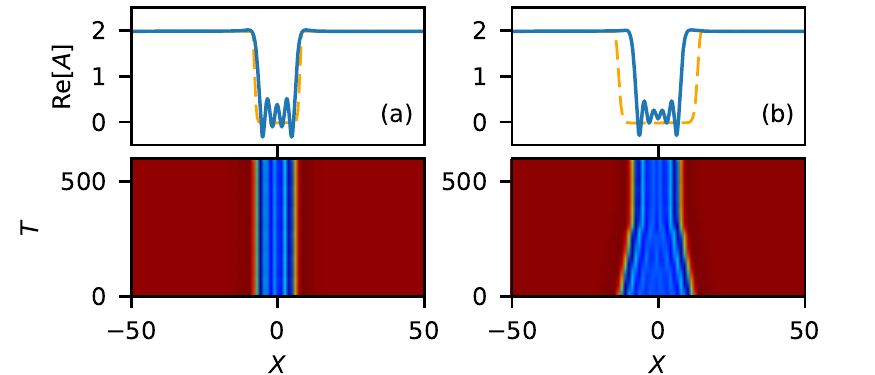}
 	\includegraphics[scale=1]{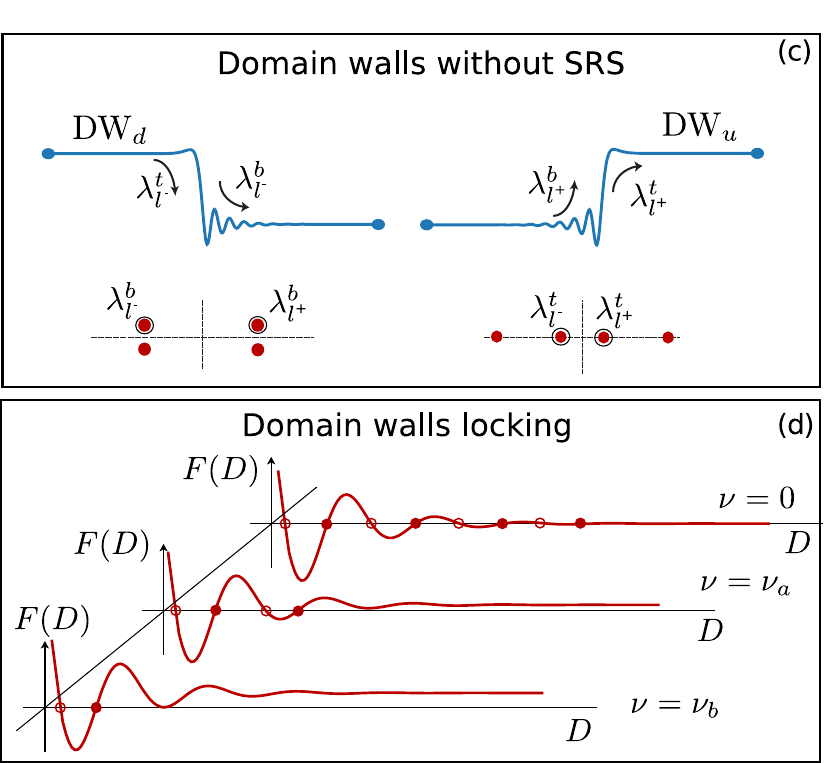}
 	\caption{Panels (a) and (b) show the formation of two different dark states starting from two distinct initial conditions. (c) shows a sketch of the profiles for DW$_u$ and DW$_d$, and the eigenvalues corresponding to each tail. In (d) we show schematically the function $F(D)$ [see Eq.~(\ref{inter})], defining the DWs locking, for three situations: at the Maxwell point ($\nu=0$), and moving apart from it $(\nu_a$ and $\nu_b>\nu_a$).  }
 	\label{fig3}
 \end{figure}

Close to the Maxwell point  of the system (e.g., for $S=2.18$ in Fig.~\ref{fig2}), two initial conditions $g^t(X)$ of different widths lead to the formation of two coexistent dark LSs like those shown in Fig.~\ref{fig3}(a)-(b).
 %This result shows initial simulation shows that close to $S_M$, LSs of distinct widths which coexist for the same range of parameters. 
%Figures~\ref{fig3} shows how the temporal evolution of two different initial profiles $g^t(X)$ [see orange dashed line in Fig.~\ref{fig3}(a)-(b)] of the form (\ref{gauss}) leads to the formation of two LSs of different widths. To perform these temporal simulations we consider a pump value $S=2.18$, very close to $S_M$.
% In Fig.~\ref{fig3}(a) the initial pulse converge very fast to a {\it dark} LS of similar width with three dips [see blue profile]. In Fig.~\ref{fig3}(b) a wider initial pulse yields, after a longer transient, to a much broader dark state.
  %From this initial test we can already infer that close to the Maxwell point DWs can lock at different separations leading to dark LSs of distinct widths which coexist for the same range of parameters.
The formation of these LSs and their coexistence is a well-understood phenomenon mediated by the interaction and locking of DWs \cite{coullet_localized_2002,coullet_nature_1987,parra-rivas_dark_2016,parra-rivas_localized_2019}. Let us briefly review this mechanism. In the absence of SRS, DW$_{u,d}$ profiles look like those shown schematically in Fig.~\ref{fig3}(c). These DWs 
exhibit {\it monotonic tails} around $A_t$, and {\it oscillatory tails} about $A_b$.
These tails can be asymptotically described by $A(X)=A_{b(t)}+\epsilon e^{\lambda X}$, where $\epsilon\ll1$ and $\lambda$ is a complex number, solution of the eigenvalue equation% \cite{parra-rivas_dark_2016}
\begin{equation}\label{biqua}
\lambda^4-(4I_h-2\Delta)\lambda^2+\Delta^2+3I_h^2-4\Delta I_h+1=0.
\end{equation}
Figure~\ref{fig3}(c) shows the set of eigenvalues associated with $A_{b}$ (left) and $A_{t}$ (right). While the eigenvalues associated with $A_{t}$ are composed by four purely real eigenvalues $(\pm Q_1,\pm Q_2)$, those associated with $A_b$ are
complex conjugates $(\pm Q\pm iK)$. 
In this context, the shape of the tails is determined by the slowest mode $e^{\lambda_iX}$, and therefore associated with the leading eigenvalue (i.e., the one with the smallest $|\lambda_i|$), that we label $\lambda_l^{t,b}$ [see $\circ$ in Fig.~\ref{fig3}(c)].
%Figure~\ref{fig3}(c) shows schematically the real part of the DW$_{u,d}$ profiles in the absence of SRS. A closer look of these states shows that, while DW$_d$ leaves $A_t$ monotonically, it approaches $A_b$ {\color{magenta} through damped oscillations}. A similar {\color{magenta} behaviour} occurs for DW$_u$. In the following we refer to the parts of the DWs far from the core as {\it tails}. Then we can say that DW$_d$ exhibits {\it monotonic tails} around $A_t$ and {\it oscillatory tails} about $A_b$.
%These tails can be asymptotically described by $A(x)=A_{b(t)}+\epsilon e^{\lambda X}$, where $\epsilon\ll1$ and $\lambda$ is a complex number, solution of the . By direct substitution of the previous ansatz in Eq.~(\ref{Eq.1}) one obtains, at first order in $\epsilon$, the eigenvalue equation \cite{parra-rivas_dark_2016}:
%\begin{equation}\label{biqua}
%\lambda^4-(4I_h-2\Delta)\lambda^2+\Delta^2+3I_h^2-4\Delta I_h+1=0,
%\end{equation}
%whose solutions correspond to the 
%set of eigenvalues $\{\lambda^{t,b}_i\}_{i=1,\dots,4}$ plotted in Fig.~\ref{fig3}(c), and associated with $A_{(t,b)}$: whereas $\{\lambda^{t}_i\}$ is composed by four purely real eigenvalues $(\pm Q_1,\pm Q_2)$, in
%$\{\lambda^{b}_i\}$ the eigenvalues are complex conjugates of the form $(\pm Q\pm iK)$. 
%The shape of the tails is then determined by the slowest mode $e^{\lambda_iX}$, and therefore is associated with the smallest (leading) $|\lambda_i|$, ($i=1,\dots,4$), which in what follows we label as $\lambda_l^{t,b}$. In Fig.~\ref{fig3}(c) we mark these eigenvalues using $\circ$.
Therefore, while the tail of DW$_d$ around $A_t$ is asymptotically described by 
%\begin{equation}
$	{\rm DW}_d(X)-A_t\sim e^{\lambda_{l^-}^tX}$,
%\end{equation}
the one around $A_b$ is described by
\begin{equation}
 {\rm DW}_d(X)-A_b\sim e^{\lambda_{l^-}^bX}=e^{-QX}{\rm cos}(KX),	
\end{equation}
with $Q={\rm Re}[\lambda_{l^-}^b]$ and $K={\rm Im}[\lambda_{l^-}^b]$.

Close to the Maxwell point $S_M$, the interaction of these DWs can be qualitatively described by the equation 
\begin{equation}\label{inter}
\partial_tD=\varrho e^{QD}{\rm cos}(KD)+\nu\equiv F(D),
\end{equation}
where $D$ is the separation between DWs, $\nu\sim S-S_M$, $\varrho$ is a positive constant depending on the system parameters \cite{coullet_nature_1987,coullet_localized_2002,clerc_analytical_2010,clerc_patterns_2005}. We have to point out that this equation cannot be explicitly derived from our model, and has been included here for illustrating the mechanism of DWs locking.

%, and $Q$ and $K$ are the real and imaginary part of the leading spatial eigenvalue describing asymptotically the tails of the DW.  
%%%%%%%%%%%%%%%%%%%%%%%%%%%%%%%%%%%%%%%
At $S_M$ ($\nu=0$) the fixed points of this equation
$D^s_n=\frac{\pi}{2K}(2n+1)$, with $n=0,1,2,\dots$,
correspond to the stationary distances at which the locking of DWs occurs \cite{coullet_nature_1987,coullet_localized_2002,clerc_patterns_2005}. Figure~\ref{fig3}(d) shows these points and their stability using $\rc{\bullet}$ for stable separations and $\rc{\circ}$ for unstable ones.  
If $S\neq S_M$, the stationary separations $D^s_n$ are slightly modified by the factor $\nu$, such that  the more we increase the separation $\nu$ from $S_M$, the less LSs form, until eventually no more locking takes place.
When the tails are monotonic ($K=0$), two DWs attract each other until they annihilate one another in a process called coarsening \cite{allen_microscopic_1979}.
%%%%%%%%%%%%%%%%%%%%%%%%%%%%%%%%%%%%%%
% Description of the interaction in our case

The DWs studied here [see Fig.~\ref{fig3}(c)] exhibit oscillatory tails around $A_b$, and are behind the formation of the dark LSs shown in Fig.~\ref{fig3}(a)-(b). In contrast, the tails around $A_t$ are monotonic, and therefore, no LSs form.

% Bifurcation structure: Collapsed snaking

\subsection{Bifurcation structure for the dark localized states: Collapsed snaking }

\begin{figure}[t]
	\centering
	\includegraphics[scale=0.95]{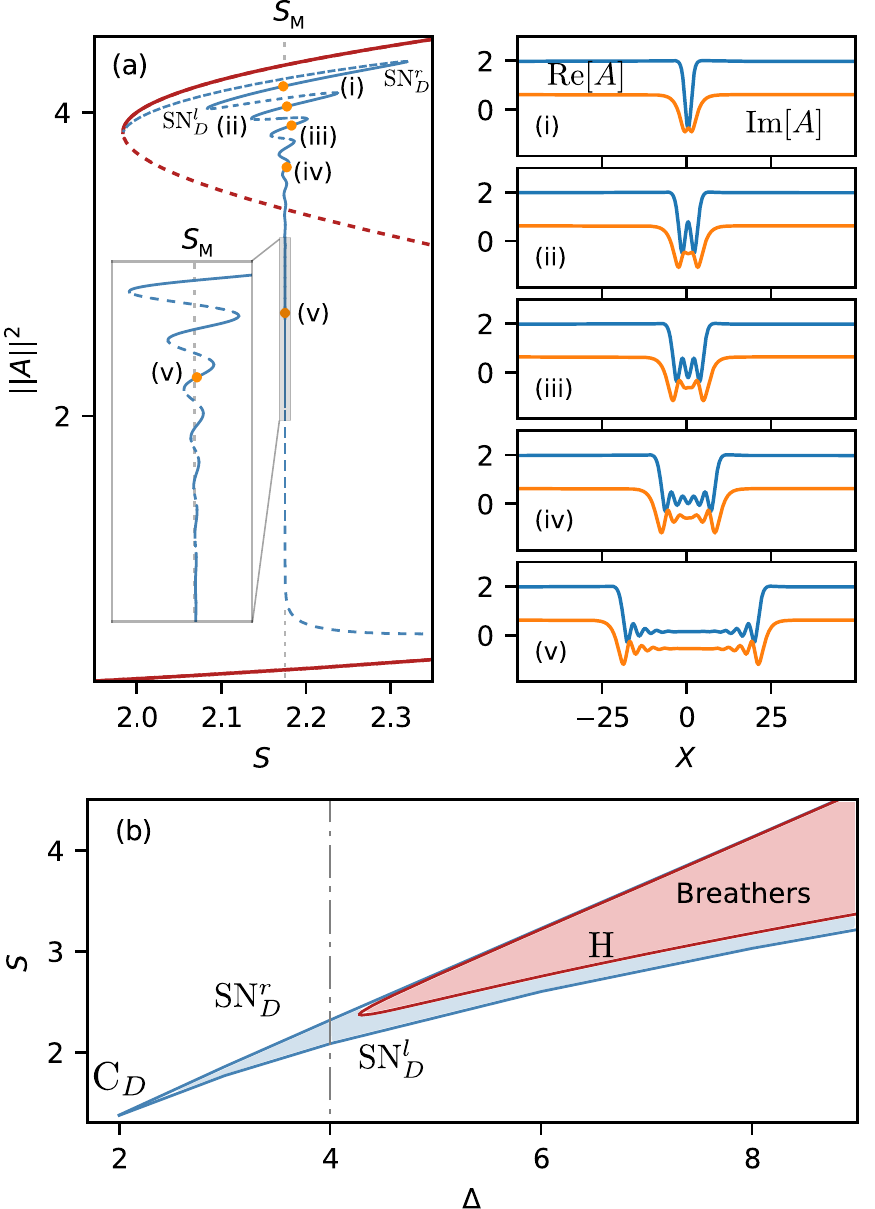}
	\caption{Panel (a) shows the collapsed snaking bifurcation diagram (blue lines) for $\Delta=4$ in the absence of SRS $(f_R=0)$. This diagram shows the norm $||A||^2$ of the different states as a function of $S$. The dark red lines represent the CW state. Stable (unstable) branches are shown with solid (dashed) lines. The labels (i)-(vi) correspond to the dark LSs shown on the right. Panel (b) shows the phase diagram in the $(\Delta,S)-$parameter space. The blue lines correspond to SN$_D^{l,r}$, and the red ones to the Hopf bifurcation H.}
	\label{fig5}
\end{figure}

The dark LSs formed through this mechanism are organized in a bifurcation diagram like the one shown in Fig.~\ref{fig5}, where we plot the $L_2$-norm 
$$||A||^2=\frac{1}{l}\int_{-l/2}^{l/2}|A(x)|^2dx,$$
of the different steady states as a function of $S$. The red lines correspond to the CW states discussed in Sec.~\ref{sec:2}, and the vertical dashed line marks the Maxwell point of the system $S_M$.  We have computed this diagram fixing $\Delta=4$, and performing a numerical parameter continuation on $S$ based on a predictor-corrector method, as described in \cite{doedel_numerical_1991,doedel_numerical_1991-1}.

The dark LSs formed through the locking of DWs undergo {\it collapsed snaking} \cite{knobloch_homoclinic_2005-1, yochelis_reciprocal_2006, parra-rivas_dark_2016,parra-rivas_localized_2019}: nearby $S_M$ the LSs solution branches (see blue lines in Fig.~\ref{fig5}) oscillate back and forth in $S$ with an amplitude which decreases as descending in $||A||^2$. The labels (i)-(vi) correspond to the dark LSs of different widths shown on the right. While decreasing 
$||A||^2$ the LSs broaden as a result of the addition of tails wavelengths, until the DWs reach the domain width. At this stage, the solution branch, accumulated around $S_M$, leaves that point and connect back to the $A_b$ CW branch at the MI (see Fig.~\ref{fig2}) \cite{parra-rivas_dark_2016}.  

The stability of these states is marked using solid (dashed) lines for stable (unstable) states, and has been obtained 
by solving the eigenvalue problem
$$\mathcal{L}\psi=\sigma\psi,$$
where $\mathcal{L}=\mathcal{L}(A)$ is the linear operator associated with the right-hand side of Eq.~(\ref{Eq.2}) evaluated at a given LS, and $\sigma$ and $\psi$ are the eigenvalues and eigenmodes associated with $\mathcal{L}$. Note that we solve this problem numerically, and therefore, $\mathcal{L}$ corresponds to the Jacobian matrix obtained from the discretization of Eq.~(\ref{Eq.2}).

This bifurcation structure follows directly from the damped oscillatory DWs interaction described by Eq.~(\ref{inter}). To understand this correspondence let us relate the sketch shown in Fig.~\ref{fig3}(d) and the collapsed snaking of Fig.~\ref{fig5}(a). At the Maxwell point $S_M$ (i.e., for $\nu=0$), a number stable and unstable dark LSs form at the stationary DWs separations $D^s_n$. Then the stable (unstable) LSs in Fig.~\ref{fig3}(d) correspond to a set of points on top of the stable (unstable) branches of solutions at $S_M$ in Fig.~\ref{fig5}(a).

When $S$ separates from $S_M$ [see Fig.~\ref{fig3}(d) for $\nu=\nu_a$] 
less stationary separations $D^s_n$ occur, resulting in the disappearance of solution branches corresponding to wider dark LSs. Increasing further the value of $S$ [see Fig.~\ref{fig3}(d) for $\nu=\nu_b$], only two intersections occur and only the stable and unstable single peak solution branches remain. Proceeding in this way eventually no more intersections take place, resulting in the complete disappearance of the LSs. Note that the SNs in the collapsed snaking occur at the tangencies shown in Fig.~\ref{fig3}(d).
For very large separations (i.e. small $||A||^2$) the interaction is very weak, and wide dark LS branches approach asymptotically the Maxwell point $S_M$. 

Dark LSs persist for different values of $\Delta$ and undergo temporal oscillatory (i.e., Hopf) instabilities that make them breathe \cite{parra-rivas_dark_2016}. Figure~\ref{fig5}(b) shows the phase diagram of these states in the $(\Delta,S)-$parameter space, where the first two folds of the dark LSs SN$_D^{l,r}$  and the Hopf instability H are plotted. Within the light red area dark LSs undergo oscillatory dynamics, whereas in the light blue one they are stable. Increasing $\Delta$, the region of existence of these regimes broadens. However, when decreasing $\Delta$ they shrink until eventually the different LSs disappear in a sequence of cusp bifurcations \cite{parra-rivas_dark_2016}. For simplicity here we only plot the cusp $C_D$ associated with the dark state shown in the profile (i).

\section{Coexistence of bright and dark localized states in the presence of stimulated Raman scattering}\label{sec:4}
%In Sec.~\ref{sec:3} we have introduced the concept of DWs locking as a mechanism for the formation of LSs, and for doing so we have considered the standard LL Eq.~(\ref{Eq.1}). We have shown that the locking occurs due to the presence of oscillatory tails in the DWs profiles, and that the later are defined by the spatial eigenvalues of CW states. 
%In this section we study how the SRS modifies the nature of the DWs tails, and therefore, the formation of LSs, allowing to not only the emergence of dark states, but also the appearance of bright LSs.  
The dynamics, interaction and locking of DWs can be strongly modified by the influence of  high-order dispersion effects \cite{parra-rivas_coexistence_2017}, or long range interactions \cite{clerc_patterns_2005,clerc_analytical_2010,escaff_non-local_2011,colet_formation_2014,gelens_formation_2014,fernandez-oto_c._strong_2014,escaff_localized_2015}, such as SRS.
% Let us briefly see some basic features of the influence of SRS on the formation of LSs. 
%The nature of the DWs tails, and therefore the formation of LSs, can be modified by the influence of other effects, such as high-order dispersion terms \cite{parra-rivas_coexistence_2017}, or other types of non-local coupling such as long range interactions such as the SRS \cite{clerc_patterns_2005,clerc_analytical_2010,escaff_non-local_2011,colet_formation_2014,gelens_formation_2014,fernandez-oto_c._strong_2014,escaff_localized_2015}.
%The SRS is an example of a phenomenon that can be modeled through 
%a long-range interaction term In our case, the SRS is modeled by  an example of the previously mentioned long range interactions, and in this section we study the implication that such effect may have on the dynamics and stability of LSs. 
\begin{figure}[t]
	\centering
	\includegraphics[scale=1]{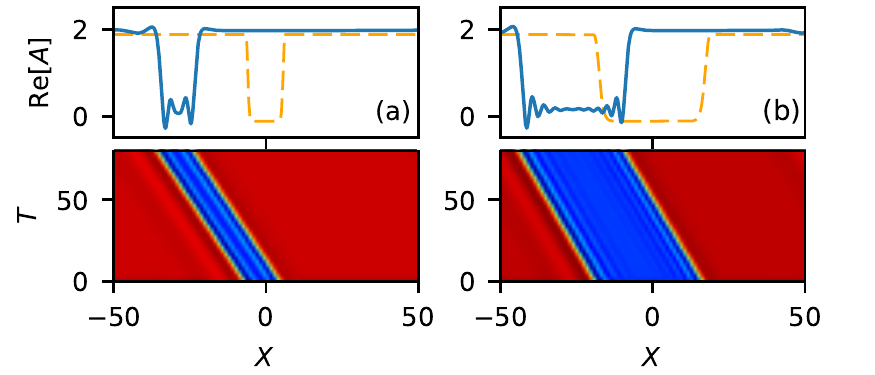}
	\includegraphics[scale=1]{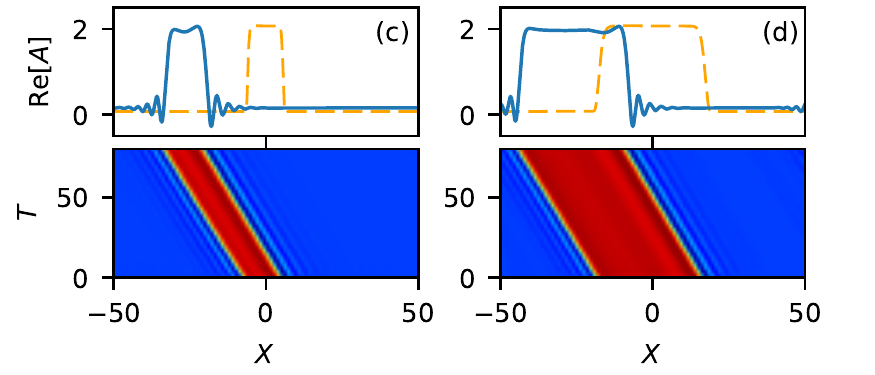}
	\includegraphics[scale=1]{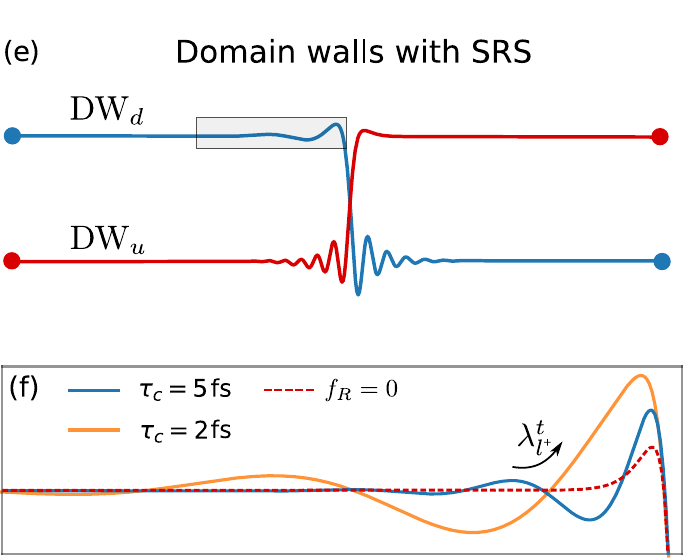}
	\caption{Formation of LSs in the presence of SRS for $\Delta=4$, $S=2.21$, and $\tau_c=5$ fs. Panels (a)-(b) show the formation and temporal evolution of dark LSs starting from two different initial conditions (see yellow dashed line). Panels (c)-(d) show the formation and evolution of bright LSs starting from different initial conditions than in (a) and (b). In (e) we plot schematically two DWs profiles corresponding to DW$_d$ in blue and DW$_u$ in red. The gray area highlights the oscillatory tails around $A_t$ plotted in detail in the close-up view of panel (f). For comparison panel (f) also shows the tails when $\tau_c=2$ fs (orange line), and in the absence of SRS (red dashed line).}
	\label{fig6}
\end{figure}
%\subsection{Modification of the oscillatory tails due to the Raman response}
Figure~\ref{fig6} shows the temporal evolution of different initial conditions of the form $g^{b,t}(X)$ in the presence of SRS.
Panels~\ref{fig6}(a) and (b) correspond to the evolution of two initial conditions $g^t(X)$ of different widths (see orange dashed profile) for $(\Delta,S)=(4,2.21)$ and $\tau_c=5$ fs. In both cases the initial condition converges fast to  asymmetric dark LSs of different widths which drift at constant speed $v$ (see blue profiles). Thus,  as for the vanishing SRS case, different type of dark LSs coexist for a single value of $S$.  
For the same parameters, similar dynamical behaviors are observed starting with initial conditions of the form $g^b(X)$, as can be seen in Figs.~\ref{fig6}(c) and (d). In this case, the system evolves to two moving bright states of different extensions, which were absent for vanishing SRS. Therefore, we can conclude that bright LSs are stabilized through SRS effect. 

The formation of these bright states can be also explained in terms of DWs interaction and locking.
%This stabilization is related with the modification of the oscillatory tails of the DWs involved in the locking process. To illustrate this let us take a look to
Figure~\ref{fig6}(e) shows the real part of DW$_d$ and DW$_u$  
for the same parameter values used in the previous temporal simulations.
Due to the SRS effect, the shape of the tails around either $A_t$ and $A_b$ differs from DW$_d$ to DW$_u$, and, as we previously stated, ${\rm DW}_{u}(X)\neq {\rm DW}_d(-X)$. A close-up view of the tails of DW$_d$ around $A_t$ [see gray box in Fig.~\ref{fig6}(e)] is shown in Fig.~\ref{fig6}(f), where we also add the tails of the unperturbed case (i.e., $f_R=0$) for comparison (see red dashed line).
While in the unperturbed case the tail of DW$_d$ leaves $A_t$ monotonically (i.e., the dominant spatial eigenvalue has the form $\lambda_{l^+}^t=+Q$), in the SRS case the tail leaves $A_t$ in a damped oscillatory manner associated with an eigenvalue $\lambda_{l^+}^t=+Q+iK$. This modification introduces a new way of interaction, such that the locking not only occurs around $A_b$, but also around $A_t$, leading to the formation of bright LSs.
In this context, the characteristic time $\tau_c$ plays an important role on the modification of the DWs tails: decreasing $\tau_c$ the wavelength of the tails increases, while its decaying weakens. This behavior is captured in  Fig.~\ref{fig6}(f) where we compare the shape of the tail for $\tau_c=5$ fs (blue line) and $\tau_c=2$ fs (orange line). Note that similar tail modifications take place in the presence of other terms breaking $X-$reflection symmetry such as third-order dispersion \cite{parra-rivas_coexistence_2017,talla_mbe_existence_2017}.

% Qualitatively, the interaction of DWs and their locking could be described by an equation similar to Eq.~(\ref{inter}) with the parameters $\varrho$, $\nu$, $Q$ and $K$ depending on the long range SRS effect.
	
Close to the nascent bistability onset, an equation describing the DWs interaction can be derived \cite{clerc_time-delayed_2020,chaos_raman}. In that limit, the resulting equation shows that the interaction and DWs locking depends on the balance between two factors: i) a contribution due to the DWs tails, and ii) a contribution directly related to the long-range interaction (see Eq.~(7) in \cite{clerc_time-delayed_2020}). One could expect that such dependence might persist in the full model (\ref{Eq.2}), with the oscillatory tails contribution to the interaction being described by Eq.~(\ref{inter}). However, the explicit derivation of an interaction equation in this context, if possible, might not be straightforward, and it is beyond the scope of the present work.

\section{Bifurcation structure for Raman dark and bright localized states}
\label{sec:5}

As we have already mentioned in Sec.~\ref{sec:2}, the collapsed snaking is determined by the damped oscillatory nature of the DWs tails through Eq.~(\ref{inter}). Therefore, any modification of the DWs tails and/or features of the interaction law, like the ones induced by the SRS, may change the LSs bifurcation structure.  

\begin{figure}[t]
	\centering
	\includegraphics[scale=0.95]{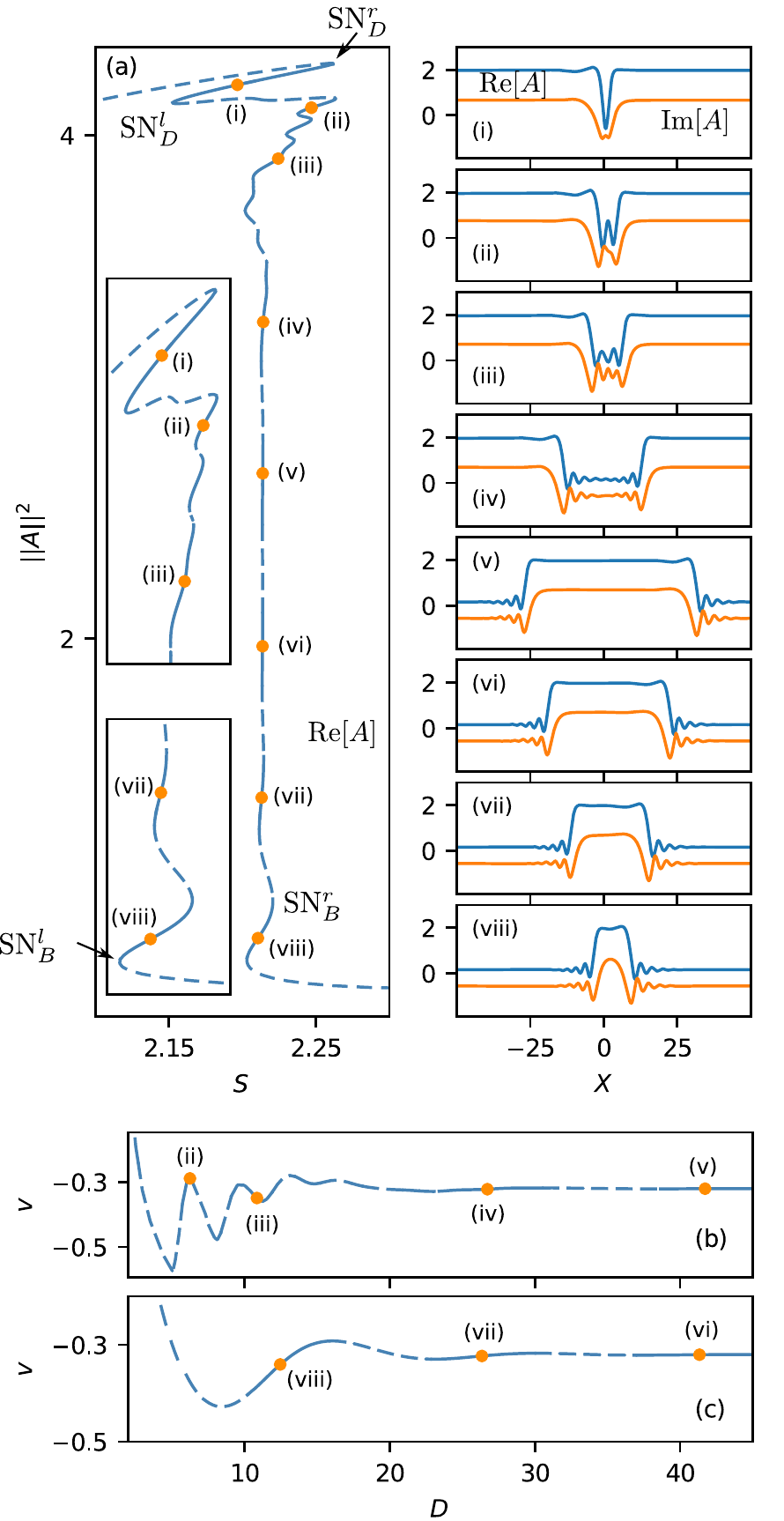}
	\caption{Panel (a) shows the LSs bifurcation diagram for $\Delta=4$ in the presence of SRS with $\tau_c=5$ fs. The labels (i)-(viii) correspond to the LSs shown on the right. Stable (unstable) solution branches are marked using solid (dashed) lines. Panel (b) shows the velocity of dark LSs as a function of their width $D$. In (c) the same than in (b) for the case of bright LSs. 
		%{\color{magenta} $[$Is it possible to use different colors for the stable and unstable variety in fig 7c and d?$]$}		%.
	}
	\label{fig7}
\end{figure}

Figure~\ref{fig7} shows the modification of the collapsed snaking diagram for $\Delta=4$ in the presence of SRS when $\tau_c=5$ fs. The labels (i)-(viii) mark the position of the LSs shown on the right. Due to the $X\rightarrow-X$ symmetry breaking these LSs drift at constant speed $v$ and are solutions of Eq.~(\ref{eq.3}). To compute these states and track them numerically in a given parameter, we need to consider a phase condition of the form $C[A]=0$ to take account for the LSs speed. Here, we define this condition as the constraint $C[A]=d{\rm Re}[A]/dX|_{X_0}=0$ which forces one extremum of the LS (maximum or minimum) to be located at $X=X_0$ \cite{parra-rivas_third-order_2014,parra-rivas_coexistence_2017}. 
In this manner, the speed of the LSs is computed as a part of the solution in the continuation algorithm. Figures~\ref{fig7}(b)-(c) show the computed speed as a function of the LSs width $D$.
The inner sub-panels in Fig.~\ref{fig7}(a) show a close-up view of the top and bottom part of the bifurcation diagram, which allows us to illustrate better the organization of those solution branches. 

The top part of the diagram corresponds to the solution branches associated with  the dark LSs presented in Sec.~\ref{sec:3}. Due to the effect of the SRS, these states are now asymmetric as shown in Fig.~\ref{fig7}(i)-(iv). 
Note that the complex form of this part of the diagram may be related to an interaction and locking process more complicated than the one described in Sec.~\ref{sec:3}. However, the confirmation of this scenario requires further investigation. 
The velocity of these states is not constant [see Fig.~\ref{fig7}(b)] but oscillates with $D$, and therefore along the diagram shown in Fig.~\ref{fig7}(a). For $D\lesssim20$ the velocity shows large oscillations with $D$ which correspond to dark states with higher $||A||^2$. For $D\gtrsim20$ (i.e., decreasing in $||A||^2$), the LSs collapse to the Maxwell point, and so does the velocity which saturates to an almost constant value.

The bottom part of the diagram shows a regular collapsed snaking in $S$, which is absent when $f_R=0$ [see Fig.~\ref{fig5}(a)]. This new bifurcation structure is related to the locking of DWs around $A_t$, which is now possible due to the presence of oscillatory tails about such state, and the presence of the long range interaction (see Sec.~\ref{sec:4})\cite{clerc_time-delayed_2020}. Four representative LS examples along this part of the diagram are shown in Fig.~\ref{fig7}(v)-(viii). These asymmetric bright LSs drift at constant speed, whose damped oscillatory dependence with $S$ is plotted Fig.~\ref{fig7}(c).

\begin{figure}[t]
	\centering
	\includegraphics[scale=0.95]{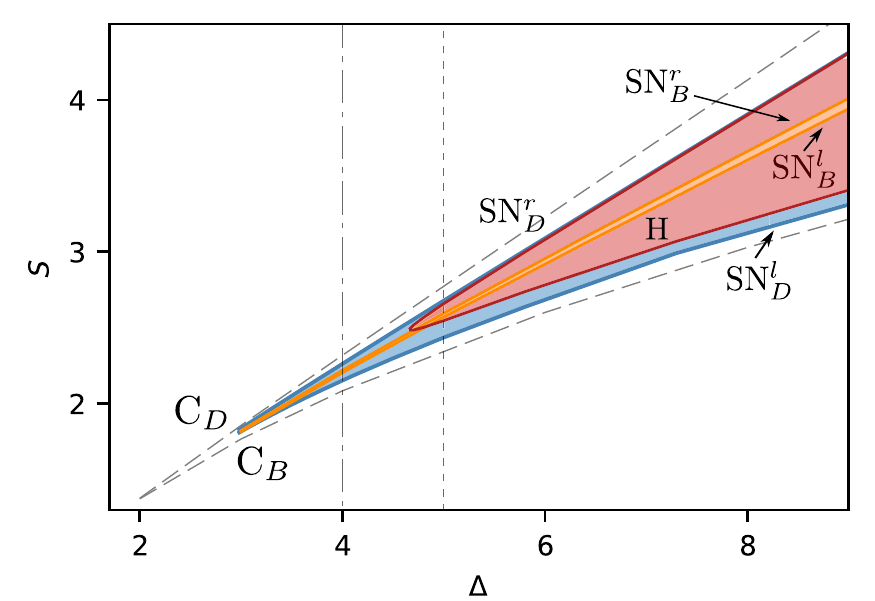}	
	\caption{Phase diagram in the $(\Delta,S)-$parameter space showing the main bifurcation lines of the system in the presence of SRS ($f_R\neq0$) for $\tau_c=5$ fs: SN$^{l,r}_D$ (blue lines) are the first two folds of the dark states, SN$^{l,r}_B$ (orange lines) are the first two folds of the bright LSs, and H (red line) corresponds to the Hopf bifurcation. The dashed gray lines correspond to SN$^{l,r}_D$ in the absence of SRS, and vertical point-dashed and dashed vertical ones to the bifurcation diagrams shown in Figs.~\ref{fig7} and \ref{fig9} respectively.}
	\label{fig8}
\end{figure}

\begin{figure}[t]
	\centering
	\includegraphics[scale=0.95]{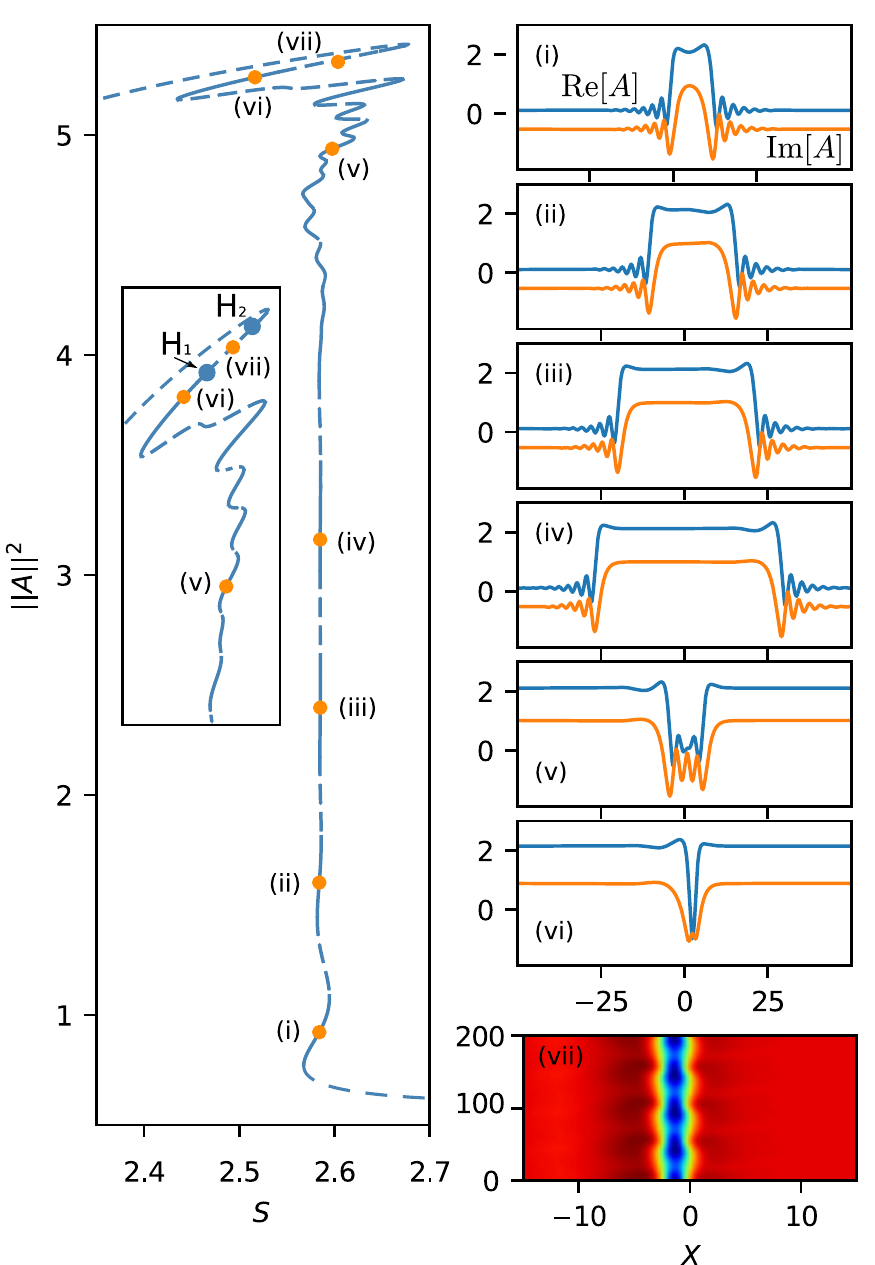}
	\caption{Collapsed snaking for $\Delta=5$ and $\tau_c=5$ fs corresponding to the dashed vertical line shown in Fig.~\ref{fig8}. The labels (i)-(vi) correspond to the LSs shown on the right, and panel (vii) shows the evolution of a breather within several oscillatory periods. }
	\label{fig9}
\end{figure}

%\subsection{Phase diagrams in the $(\Delta,S)-$parameter space}
Dark and bright LSs persist for different values of $\Delta$ as shown in the phase diagram of Figure~\ref{fig8}, where we show the main bifurcation lines of the system. This diagram has been computed through a two-parameter continuation in $(\Delta,S)$ while fixing $\tau_c=5$ fs. 
As in Fig.~\ref{fig5}(b) the blue lines correspond to the first two folds of the single dip dark LS SN$_{D}^{l,r}$, the red line is the Hopf bifurcation undergone by these states, and in orange we plot the first two folds of the bright LSs, namely SN$_{B}^{l,r}$ [see Fig.~\ref{fig7}(a)]. The dashed gray lines correspond to SN$_{D}^{l,r}$ in the absence of SRS (i.e., $f_R=0$), and have been added for comparison.
The vertical pointed-dashed line corresponds to the bifurcation diagram shown in Fig.~\ref{fig7} for $\Delta=4$, and the dashed one to Fig.~\ref{fig9}.

In the presence of SRS, the different dynamical regions shown in Fig.~\ref{fig5}(b), such as the region of existence of single dip dark LSs and the breathing region shrink, leading in this way to a partial stabilization of the previous breathing dark states. %\bc{ Note that this behavior is similar to the one accounted in the context of anomalous ....}
The area in-between SN$_{B}^{l,r}$ (see light orange area) shows the region of existence of the bright LSs, which widens increasing $\Delta$.
Decreasing $\Delta$, however, this region shrinks until eventually SN$_{D}^{l}$ and SN$_{D}^{r}$ collide and disappear in a cusp bifurcation $C_B$, which occurs approximately for the same values than $C_D$.

Figure~\ref{fig9} shows the bifurcation diagram for $\Delta=5$. For this value of $\Delta$ the bottom of the diagram is very much alike the one shown in Fig.~\ref{fig7} for $\Delta=4$, and some representative examples of bright LSs are shown in panels~\ref{fig9}(i)-(iv).
The top part of the diagram, however, although morphologically similar to the one depicted in Fig.~\ref{fig7}, undergoes a Hopf instability (see close-up view in the inset of Fig.~\ref{fig9}), that makes the single dip dark state [see profile (vi)] breathe as shown in Fig.~\ref{fig9}(vii).

The collapsed snaking bifurcation structure, and the instabilities undergone by their underlying LSs persist for higher values of $\Delta$ as shown in the phase diagram of Fig.~\ref{fig8}. Note that bright LSs may also undergo oscillatory instabilities \cite{yao_generation_2020}. However, for the range of parameters studied in this work we have not observed such type of dynamics.

\section{Influence of $\tau_c$ on the localized states dynamics and stability}\label{sec:6}

\begin{figure}[t]
	\centering
	\includegraphics[scale=0.95]{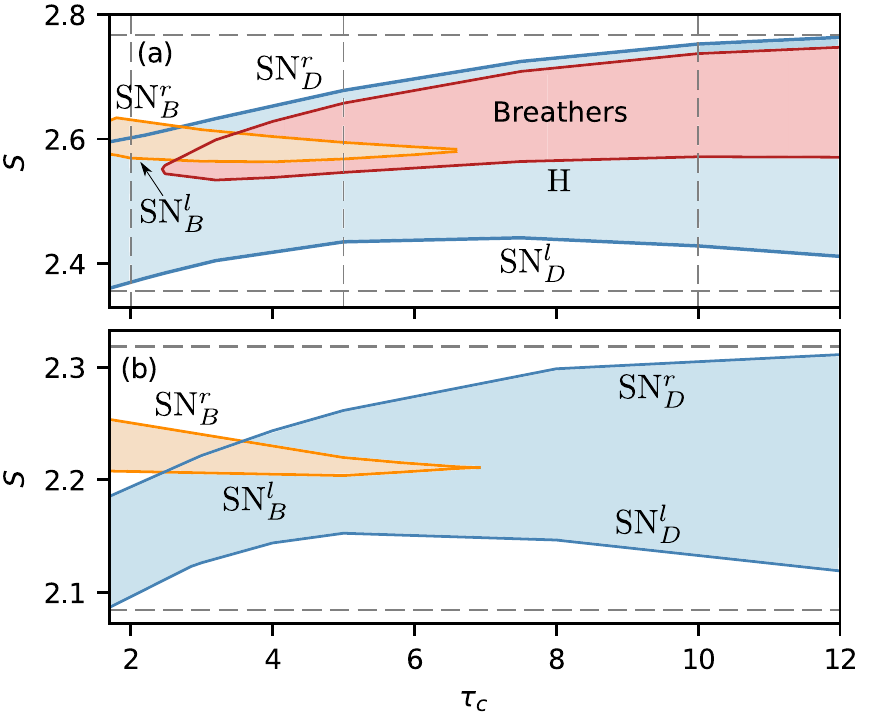}	
	\caption{Panel (a) shows the $(\tau_c,S)-$phase diagram for $\Delta=5$, where the main bifurcations of the system are plotted: SN$^{l,r}_D$ (blue lines) are the first two fold of the dark states, SN$^{l,r}_B$ (orange lines) are the first two folds of the bright LSs, and H (red line) corresponds to the Hopf bifurcation. The vertical lines in panel (a) corresponds to the bifurcation diagrams shown in Fig.~\ref{fig10}. Panel (b) shows the same type of phase diagram for $\Delta=4$. In both panels horizontal dashed lines represent the position of SN$^{l,r}_D$ for vanishing SRS.}
	\label{fig8b}
\end{figure}

So far we have studied the influence of the SRS on the dynamics and stability of LSs for a single value of the characteristic time $\tau_c=5$ fs. This parameter strongly impacts the stability of LSs as shown in the context of anomalous dispersion \cite{wang_stimulated_2018}, and one may wonder how the previous scenario modifies when varying its value. 

To clarify this point we perform a two-parameter continuation of the main bifurcations of the system in $S$ and $\tau_c$ by fixing $\Delta$. The outcome of these computations is shown in Fig.~\ref{fig8b} for two different values of detuning $\Delta$.
In Fig.~\ref{fig8b}(a) we show the $(\tau_c,S)-$phase diagram for $\Delta=5$, and
% that has been originally tracked from the section of Fig.~\ref{fig8} marked with a vertical point-dashed line.
Fig.~\ref{fig8b}(b) shows the one for $\Delta=4$. In both cases the horizontal dashed lines mark the position of SN$_{D}^{l,r}$ in the absence of SRS for comparison. Note that the Hopf bifurcation H is present for $\Delta=5$ [see red line line in Fig.~\ref{fig8b}(a)], while absent for $\Delta=4$ [see Fig.~\ref{fig8b}(b)].

\begin{figure}[t]
 	\centering
 	\includegraphics[scale=0.95]{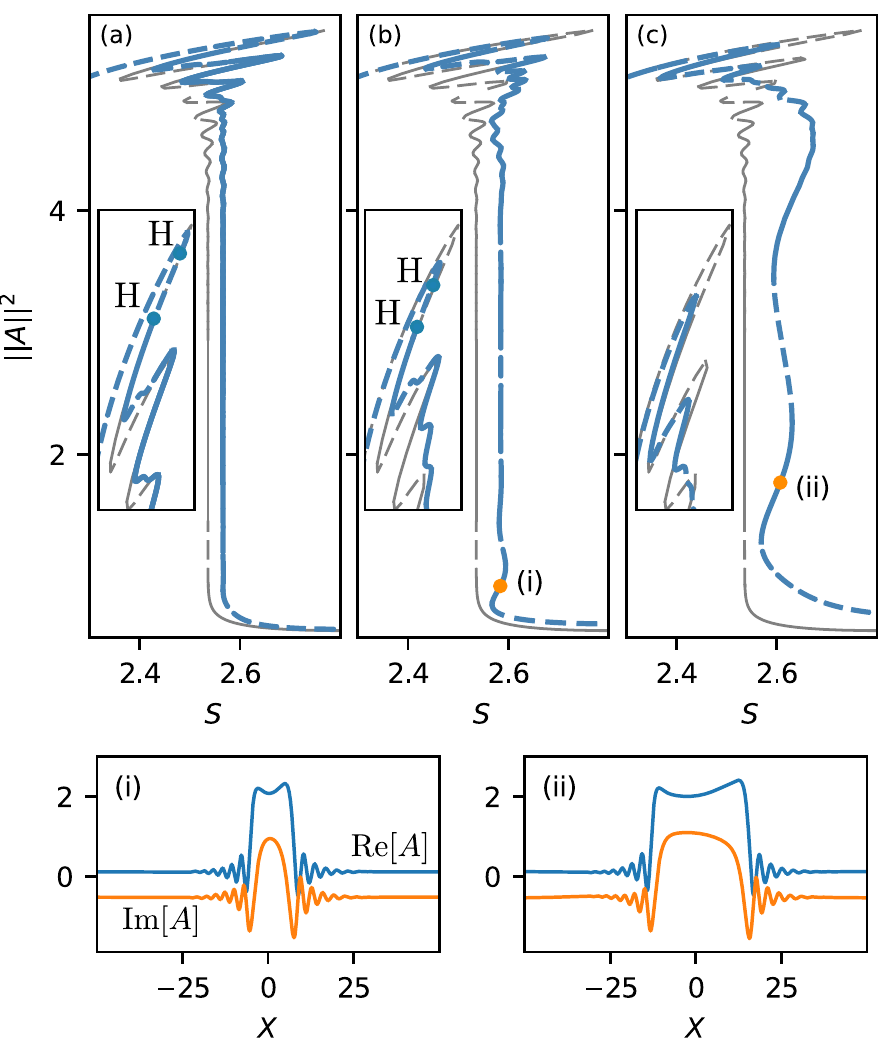}
 	\caption{Modification of the collapsed snaking structure for $\Delta=5$ while increasing $\tau_c$ (see blue diagram): in (a) $\tau_c=10$ fs, in (b) $\tau_c=5$ fs, and in (c) $\tau_c=2$ fs. In gray we have also plotted the diagram in the absence of SRS for comparison. The inset panels show a close-up view of the top part of the collapsed snaking where the Hopf instabilities H are signaled. Two examples of bright LSs profiles are plotted in sub-panels (i) and (ii).}
 	\label{fig10}
 \end{figure}

For $\Delta=5$, the modification of the collapsed snaking structure with $\tau_c$ is depicted in Fig.~\ref{fig10} for three particular values of $\tau_c$ corresponding to the vertical dashed lines shown in Fig.~\ref{fig8b}(a). 
%For $\tau_c=10{\rm fs}$ [see Fig.~\ref{fig10}(a)] the scenario is very similar to the unperturbed case (see gray diagram computed for $f_R=0$): the Hopf instability slightly modifies its position [see close-up view in Fig.~\ref{fig10}(a)], and the collapsed snaking corresponding to the bottom of the diagram is almost absent. Increasing further $\tau_c$, the Raman response function $\mathsf{R}$ becomes sharper and highly damped [see Sec.~\ref{sec:1}] tending to an almost instantaneous response, and therefore, the deviation from the unperturbed case is almost negligible.
For $\tau_c=10$ fs [see Fig.~\ref{fig10}(a)] the scenario is very similar to the unperturbed case (see gray diagram computed for $f_R=0$), although the Hopf instability slightly modifies its position as shown in the close-up view of Fig.~\ref{fig10}(a). In addition, the collapsed snaking corresponding to the bottom of the diagram is
strongly compressed. Increasing $\tau_c$ further, the Raman response function $\mathsf{R}$ becomes sharper and highly damped (see Sec.~\ref{sec:1}) tending to an almost instantaneous response, and therefore, the deviation from the unperturbed case is almost negligible.

Reducing $\tau_c$, however, the SRS modifies strongly the DWs tails (see Sec.~\ref{sec:4}), and as a consequence the LSs bifurcation structure. 
%and its effect on the bifurcation structure becomes more relevant.
Thus, while the region of existence of dark LSs and breathers shrinks, the region of existence of bright LSs broadens [see phase diagram in Fig.~\ref{fig8b}(a)]. An example of this situation is shown in Fig.~\ref{fig10}(b) for $\tau_c=5$ fs, where whereas the top part of the diagram is highly modified (see close-up view), the bottom one shows larger damped oscillations in $S$ as a result of the appearance of bright LSs like the one shown in Fig.~\ref{fig10}(i).

Decreasing  $\tau_c$ even further, the Hopf bifurcation disappears, and with it, the breathing behavior [see Fig.~\ref{fig8b}(a)], leading to the stabilization of the single dip dark states.  In this regime the bifurcation structure is similar to the one depicted in Fig.~\ref{fig10}(c) for $\tau_c=2$ fs.  For this value of $\tau_c$ the branches corresponding to the dark states are strongly modified [see the detailed view in the inset of panel~\ref{fig10}(c)], and the bright LSs increase their region of existence due to the dominant interaction of the DWs tails around $A_t$. An example of this type of bright LSs is shown in panel~\ref{fig10}(ii). For $\Delta=4$ [see Fig.~\ref{fig8b}(b)] the scenario is quite similar to the previous one, despite of the absence of the breather regime. 

%Note that while in the anomalous regime the increase of $\tau_c$ induces the stabilization of breathers \cite{wang_stimulated_2018}, here we find the opposite situation, where such stabilization occurs when decreasing $\tau_c$. 

\section{Discussion and Conclusions}\label{sec:7}
We have presented a detailed theoretical study of the dynamics and bifurcation structure of dissipative LSs emerging in externally driven Kerr cavities in the presence of stimulated Raman scattering. To perform this study we have considered the modified Lugiato-Lefever equation with the Raman response, and we have focused on the normal group velocity dispersion regime (see Sec.~\ref{sec:1}).

In the absence of SRS, the typical LSs arising in this regime are dark. These type of LSs form due to the locking of DWs which exist within a region of bi-stability between two different CW states. The locking occurs through the overlapping of the DWs tails leading to LSs of different extensions that can be seen as a portion of one CW state embedded on the other one (see Sec.~\ref{sec:2} and \ref{sec:3}).
From a bifurcation perspective and a fixed value of $\Delta$, these states undergo {\it collapsed snaking} (see Sec.~\ref{sec:3}): the LSs solution branches experience a sequence of exponentially decaying oscillations in the pump $S$ around the Maxwell point of the system, as a result of the DWs interaction and locking \cite{parra-rivas_dark_2016,parra-rivas_origin_2016,parra-rivas_localized_2019}. For large values of detuning $\Delta$, moreover, these states undergo oscillatory instabilities that make them breathe \cite{parra-rivas_dark_2016,parra-rivas_origin_2016}.

The presence of SRS strongly modifies the dynamics and stability of the previous states, and furthermore induces the emergence of bright LSs \cite{clerc_time-delayed_2020}.  From a dynamical point of view, the SRS effect has two main implications. First, the SRS term breaks the reflection symmetry $X\rightarrow-X$, inducing a constant drift in the, otherwise static, LSs.
% Note that in general, the drift can be induced in many other systems by odd chromatic dispersion effects \cite{vladimirov_effect_2018,parra-rivas_third-order_2014,parra-rivas_coexistence_2017}, phase gradients \cite{turaev_chaotic_2008}, Ising-Bloch transitions \cite{coullet_breaking_1990,gomila_theory_2015}, external delay feedback \cite{tlidi_spontaneous_2009,panajotov_impact_2016}, and walk-off in the context of quadratic cavities \cite{zambrini_convection-induced_2005}. 
Second, the SRS modifies the spatial eigenvalues of the CW states (i.e., the shape of the DWs oscillatory tails), and the interaction and locking of DWs. As a result of this alteration, the dark states modify their dynamics, and bright LSs arise (see Sec.~\ref{sec:4}). 

Note that the drift can be also induced by other mechanisms such as odd chromatic dispersion effects \cite{tlidi_drift_2013,parra-rivas_third-order_2014,parra-rivas_coexistence_2017,vladimirov_effect_2018}, phase gradients \cite{turaev_chaotic_2008}, and external delay feedback \cite{tlidi_spontaneous_2009,panajotov_impact_2016}, among others  \cite{coullet_breaking_1990,gomila_theory_2015,zambrini_convection-induced_2005}. 

Dark and bright Raman LSs undergo collapsed snaking that we have characterized in detail as a function of $\Delta$ and $S$ for a fixed value of the characteristic time parameter $\tau_c$.
Figure~\ref{fig8} in Sec.~\ref{sec:5} summarizes the main dynamical regimes in the $(\Delta,S)-$phase space for $\tau=5$ fs. As shown in \cite{wang_stimulated_2018,clerc_time-delayed_2020}, the parameter $\tau_c$ has important implications regarding the LSs stability. These implications have been analyzed in detail in Sec.~\ref{sec:6}, and the main results are summarized in Fig.~\ref{fig8b}.  The larger the value of $\tau_c$, the stronger the SRS effect and modification of the DW tails, resulting in a more complex collapsed snaking structure (see Fig.~\ref{fig10}). Note that $\tau_c=\sqrt{L|\beta_2|/2\alpha}$, and we can reinterpret the previous results in terms of the cavity length $L$, losses $\alpha$, and chromatic dispersion coefficient $\beta_2$. Thus, if we fix $L$ and $\alpha$, $\tau_c\propto \sqrt{|\beta_2|}$, and we can reformulate the conclusion of Sec.~\ref{sec:6} in terms of $\beta_2$. Therefore, whereas increasing $\beta_2$ the existence region of dark LSs and breathers widens, the one of bright LSs shrinks, and eventually disappears.

A similar scenario, also supported experimentally, can be found when the $X-$reflection symmetry is broken through third order chromatic dispersion \cite{parra-rivas_coexistence_2017, li_experimental_2020}. In such case, the modification of the DWs tails about $A_b$ is much more prominent than in the presence of SRS, and as a consequence, the existence region of bright LSs is much wider. Furthermore, the extension of this region increases with the strength of the third order dispersion, in contrast to the SRS case where it decreases with $\tau_c$. %In principle, similar scenarios may be found for any other physical effects breaking the $X-$reversibility and } 
%-----------------------------------------------------\\

%This work has been performed for a fixed domain width $l=100$. However, different extensions of the domain may influence the number of different LSs appearing in the cavity as the wavelength of the DWs tails becomes comparable to the domain size. Thus, the larger the domain, the wider the states that can emerge in the system.  

This work has been performed for a fixed domain width $l=100$, although the results presented here can be generalized to different domain extensions. As the LSs width is an integer multiple of the DWs tails wavelength, the size of the domain can strongly constraint the
variety of LSs allowed in the system. Thus, the larger the domain (i.e., the cavity), the wider the states that can emerge in the system.  Some physical parameter values of all fiber cavities, for which the observation of these type of structures may be possible, are presented in \cite{clerc_time-delayed_2020}.
%-----------------------------------------------------\\

In conclusion, we have shown that SRS strongly impacts the LSs formation, dynamics, stability, and bifurcation structure in the bi-stability scenario typical of the normal dispersion regime. The modifications and features of this scenario have been analyzed and characterized in detail. We have shown that Raman dark and bright LSs are robust, and  persist under modification of different control parameters of the system.

%With this theoretical work we expect to guide experiments in this type of systems, as has been done successfully in the past \cite{garbin_experimental_2017,li_experimental_2020}.

 %The final aim of this type of work is to complete previous results in the topic, and we believe that will result useful for guiding experiments in this type of systems as has been done previously \cite{garbin_experimental_2017,li_experimental_2020}.

The final aim of studies of this kind is to be useful for understanding 
the formation and dynamics of LSs in Kerr nonlinear optical cavities, and
guiding experiments in this type of systems, as has been done previously \cite{garbin_experimental_2017,li_experimental_2020}.

%In conclusion, we have shown that SRS 
%strongly impacts the formation, dynamics, stability, and bifurcation structure bright and dark LSs emerging in the normal dispersion regime.

%We have shown that Raman LSs are robust, and  persist under modification of different control parameters of the system.  The final aim of this type of work is to guide experiments in this type of systems as has been done previously \cite{garbin_experimental_2017,li_experimental_2020}.

\acknowledgments 
PPR and MT acknowledge support from 
the Fonds National de la Recherche Scientifique F.R.S.-FNRS, (Belgium). SC acknowledges the LABEX CEMPI (ANR-11-LABX-0007) as well as the Ministry of Higher Education and Research, Hauts de France council and European Regional Development Fund (ERDF) through the Contract de Projets Etat-Region (CPER Photonics for Society P4S). MC acknowledges funding from Millennium Institute for Research
in Optics (MIRO) and FONDECYT projects 1180903.

\bibliographystyle{ieeetr}
\bibliography{Ramanbiblio}

\end{document}
\section{The Raman term}

\begin{multline}
\mathcal{F}[\mathsf{R}]=\int_{-\infty}^{\infty}\mathsf{R}(x)e^{ikx}dx=
\int_{0}^{\infty}R'(x)e^{ikx}dx=\\\eta \int_{0}^{\infty}e^{-\alpha x}{\rm sin}(\beta x)e^{ikx}dx=\frac{\eta}{2i} \int_{0}^{\infty}e^{(ik-\alpha)x}(e^{i\beta x}-e^{-i\beta x})dx=\\\\
\frac{\eta}{2i} \int_{0}^{\infty}e^{[i(k+\beta)-\alpha]x}dx-\frac{\eta}{2i} \int_{0}^{\infty}e^{[i(k-\beta)-\alpha]x}dx=\\\\\frac{\eta}{2}\left(\frac{1}{(k+\beta)+i\alpha}-\frac{1}{(k-\beta)+i\alpha}\right)=\\\\
\frac{-\beta \eta}{\left[(k+\beta)+i\alpha\right]\left[(k-\beta)+i\alpha\right]}=\\\\\frac{\beta\eta(\alpha^2+\beta^2-k^2)}{(\alpha^2+\beta^2-k^2)^2+4\alpha^2k^2}+i\frac{2\beta\eta\alpha k}{(\alpha^2+\beta^2-k^2)^2+4\alpha^2k^2}
\end{multline}

\appendix